\documentclass[12pt,a4paper]{article}
\usepackage{graphicx}
\usepackage{color}
\usepackage{citesort}
\usepackage{latexsym}
\usepackage{amsmath}
\usepackage{amssymb}

\newcommand{\ie}{\textit{i.e.}}
\newcommand{\eg}{\textit{e.g.}}
\newcommand{\taud}{\tau_{\rm d}}
\newcommand{\vthr}{v_{\rm thr}}

\begin{document}

\title{Bursty communication patterns facilitate spreading\\ in a threshold-based epidemic dynamics}
	
\author{Taro Takaguchi$^1$, Naoki Masuda$^{1}$, and Petter Holme$^{2,3,4,*}$
\\
\\
\\
${}^{1}$ 
Department of Mathematical Informatics,\\
The University of Tokyo,\\
7-3-1 Hongo, Bunkyo, Tokyo 113-8656, Japan\\
\\
${}^{2}$
IceLab, Department of Physics, Ume\r{a} University,\\ 
901 87 Ume\r{a}, Sweden\\
\\
${}^{3}$
Department of Energy Science, Sungkyunkwan University,\\
Suwon 440-746, Korea\\
\\
${}^{4}$
Department of Sociology, Stockholm University,\\
106 91 Stockholm, Sweden\\
\\
* Corresponding author (petter.holme@physics.umu.se)
}

\setlength{\baselineskip}{0.77cm}
\maketitle

\newpage

\begin{abstract}
\setlength{\baselineskip}{0.77cm}
Records of social interactions provide us with new sources of data for understanding how interaction patterns affect collective dynamics.
Such human activity patterns are often bursty, \ie, they consist of short periods of intense activity followed by long periods of silence.
This burstiness has been shown to affect spreading phenomena; it accelerates epidemic spreading in some cases and slows it down in other cases.
We investigate a model of history-dependent contagion.
In our model, repeated interactions between susceptible and infected individuals in a short period of time is needed for
a susceptible individual to contract infection. 
We carry out numerical simulations on real temporal network data to find that bursty activity patterns facilitate epidemic spreading in our model. 
\end{abstract}

\newpage

%%%%%%%%%%%%%%%%%%%%%%%%%%%%
\section{Introduction}
Communication between individuals is a fundament of human society.
Nowadays technologies such as sensor devices and online communication services provide us
with records of interaction between individuals, including face-to-face conversations, e-mail exchanges, and phone calls,
in massive amounts.
Such data often consist of a sequence of interaction events.
Each event is represented by a triplet, \ie, the IDs of two individuals involved in the event and the time of the event.
One traditional way to characterize such data is to represent them as an aggregated network, in which the links are drawn between two nodes (\ie, individuals) that communicate in at least one event,
and investigate structural properties of the aggregated static networks~\cite{Newman2010}.
Another and richer representation of this type of data is to model them as temporal networks, in which the links between two nodes exist only at the time of an event~\cite{Holme2011}.

Effects of temporal networks on contagious phenomena, 
such as infectious diseases and rumors, have been investigated by various authors. 
To simulate spreading dynamics on temporal networks, we read the events in an empirical event sequence one by one in the chronological order
and possibly update the states (\eg, susceptible and infected) of the two nodes involved in the event.
Karsai and colleagues simulated the susceptible-infected (SI) model on temporal networks and found that bursty activity patterns slow down contagions~\cite{Karsai2011};
Bursty activity patterns are identified with a long-tailed distribution of the interevent intervals (IEIs)~\cite{Barabasi2005,AVazquez2006}.
The slowing down occurs because, at an arbitrary time point,
the average time to the next event is longer for the long-tailed IEI distribution than for the exponential IEI distribution with the same mean.
In other words, after an individual gets infected, it tends to take longer time to infect the neighbors under the long-tailed as compared to exponential IEI distribution.
Other numerical~\cite{Miritello2011,Stehle2011} and analytical~\cite{AVazquez2007,Iribarren2009,Karrer2010} results
also support that the long-tailed IEI distribution mitigates contagion.
However, the burstiness was reported to accelerate contagion on a different data set~\cite{Rocha2011} and a different type of  epidemic dynamics~\cite{Karimi2012}.
Our understanding of the effect of the burstiness on contagious processes is still elusive.

In the present study, we show that bursty activity patterns facilitate epidemic spreading in a variant of the deterministic threshold model~\cite{Dodds2004,Dodds2004-1}.
In standard models of epidemics including the SI, susceptible-infected-recovered (SIR), and susceptible-exposed-infected-recovered (SEIR) models, which have been employed in the literature cited above,
a susceptible node gets infected from an infected neighbor with a constant probability in an event,
regardless of the amount of exposure to infected neighbors in the past.
However, history-dependent thresholding effects in which the thresholding operates on the concentration of the pathogen
have been reported for some infectious diseases mediated by bacteria, such as the tuberculosis and the dysentery~\cite{Joh2009}.
In the case of information propagation, the exposure to the information increases one's interest in a topic,
and the attractiveness of a topic decays in time in the absence of stimulus~\cite{Crane2008,Averell2011}.
We may need multiple interactions to persuade others to do something,
and repeated contacts in a short period can be more effective than those dispersed over a long period.
To consider this type of infection, we generalize the deterministic threshold model to the case of history dependence and memory decay and simulate the proposed model on temporal network data.

%%%%%%%%%%%%%%%%%%%%%%%%%%%%%%%%%%%%%%%%%%%%
\section{Methods}
Each node $i$ is assumed to have an internal variable denoted by $v_i \geq 0$ ($i=1,2,\ldots,N$),
which represents, for example, the concentration of a pathogen in the individual or the individual's interest in a topic.
Initially, $v_i$ to equal to zero for all $i$. 
We assume that node $i$ is in the susceptible ($S$) state before $v_i$ exceeds a threshold value $\vthr$ 
and that node $i$ is in the infected ($I$) state once $v_i$ exceeds $\vthr$.
Each node is in either state.
Nodes in state $I$ never return to state $S$; our model is an extension of the SI model.
Therefore, the number of $I$ nodes monotonically increases in time.

\begin{figure}
\centering
\includegraphics[width=\hsize, clip]{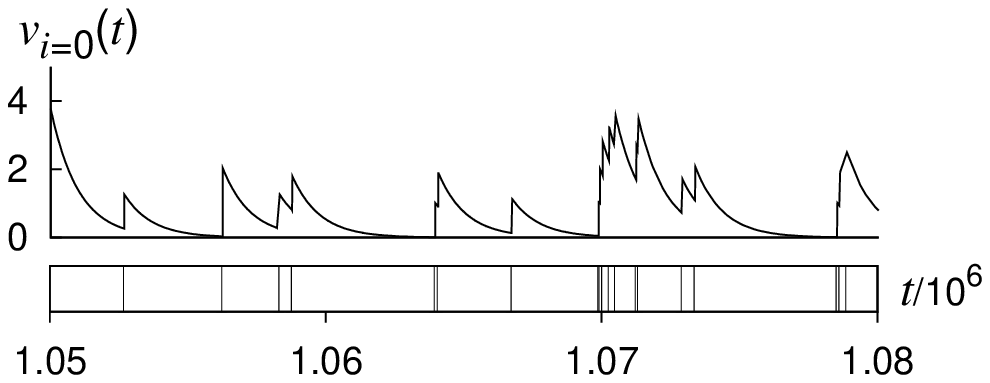}
\caption{$v_{i=0}(t)$ for $1.05\times 10^6 \leq t \leq 1.08\times 10^6$ in Email data set.
We set $\taud=1000$.
The vertical ticks in the box plot in the bottom indicate the times of the events that involve node $i=0$.}
\label{fig:v_i}
\end{figure}

When node $i$ in state $S$ interacts with an $I$ node through an event, $v_i$ is increased by unity.
In the absence of interaction with $I$ nodes, $v_i$ is assumed to decay exponentially in time.
In other words, $v_i$ is given by
\begin{equation}
v_i(t) = \sum_{t_e} \exp \left( -\frac{t-t_e}{\taud} \right),
\end{equation}
where $t_e$ is the time of an event between node $i$ and an $I$ node, and $\taud$ is the decay time constant.
An example time course of $v_i(t)$ is shown in Fig.~\ref{fig:v_i}.

The model contains two parameters $\taud$ and $\vthr$
and can be regarded as a variant of the deterministic threshold model~\cite{Dodds2004,Dodds2004-1}.
Although we assume that all the nodes have the same values of $\taud$ and $\vthr$ for simplicity,
it is straightforward to generalize the model in the case of heterogeneous parameter values.

We simulate our model numerically on empirical temporal networks in the following way.
At $t=0$, we select a node as initial seed and set its state to $I$.
All the other nodes are initially in state $S$.
Then, we chronologically read the event sequence one by one
and update $v_i$ and the states of the two nodes involved in the event.
Because our model is deterministic, the final infection size
(\ie, fraction of $I$ nodes at time $t_{\max}$, where $t_{\max}$ is the time of the last event in the data set),
denoted by $I_i$, is unique for given initial seed $i$, $\taud$, and $\vthr$.

We use two data sets.
The first data set, called Conference in the following, is the face-to-face conversation log between attendees of a scientific conference~\cite{Isella2011}.
The second data set, called Email, is the record of e-mail exchanges between the members of a university~\cite{Eckmann2004}.
In the second data set, we neglect the direction of the interaction (\ie, from sender to receiver) for simplicity.
The basic statistics of the data sets are summarized in Tab.~\ref{tab:datasets}.

%%%%%%%%%%%%%%%%%%%%%%%%%%%%%%%%%%%%%%%%%%%%%%%%
\section{Results}
\begin{figure}
\includegraphics[width=0.48\hsize, clip]{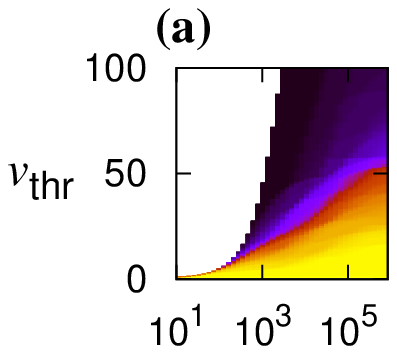}
\includegraphics[width=0.48\hsize, clip]{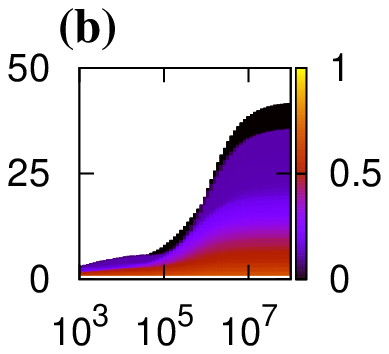}\\
\includegraphics[width=0.48\hsize, clip]{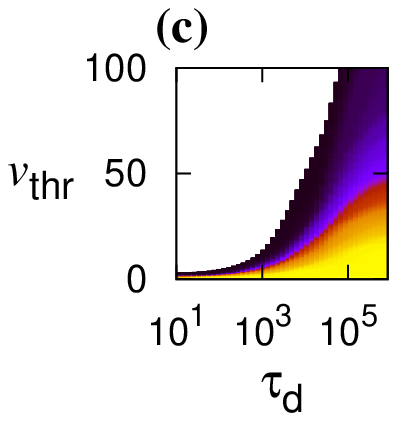}
\includegraphics[width=0.48\hsize, clip]{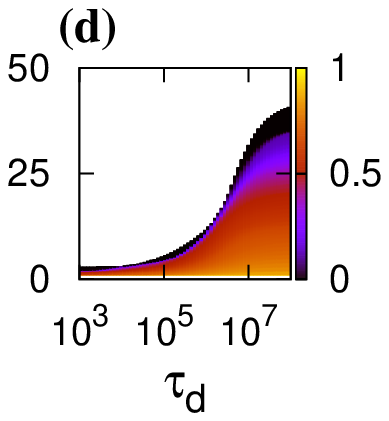}
\caption{Dependence of the final infection size $I_m$ on $\taud$ and $\vthr$.
(a), (b) Original temporal networks. (c), (d) Randomized temporal networks.
(a), (c) $I_{m=55}$ in Conference data set.
(b), (d) $I_{m=3024}$ in Email data set.
No infection occurs in the black parameter regions.
The parameter values for which at least one infection occurs are colored.}
\label{fig:I_i}
\end{figure}

In Figs.~\ref{fig:I_i}(a) and \ref{fig:I_i}(b), we plot the dependence of final infection size $I_m$ on $\taud$ and $\vthr$ for initial seed node $m$
having the maximum number of events in Conference and Email data sets, respectively.
In the blank parameter region, no infection occurs such that $I_m = 1/N$.
Naturally, $I_m$ increases with $\taud$ and decreases with $\vthr$.

Next, we carry out the same set of simulations on the randomized temporal networks for the sake of comparison.
To this end, we use the so-called randomly-permuted-times randomization,
in which the time stamps of all the events are randomly shuffled~\cite{Karsai2011,Miritello2011,Holme2011}.
The randomization eliminates temporal properties of the original temporal networks such as bursty activity patterns, daily and weekly patterns, and the pairwise correlations of the IEIs,
whereas it conserves all the properties of the aggregated networks, \ie, weighted adjacency matrix.

For the randomized temporal networks, the dependence of $I_m$ on $\taud$ and $\vthr$ are shown in Figs.~\ref{fig:I_i}(c) and \ref{fig:I_i}(d) for Conference and Email data sets, respectively.
We find that the parameter region in which infection occurs is larger for the original temporal networks (colored regions in Figs.~\ref{fig:I_i}(a) and \ref{fig:I_i}(b)) than for the randomized temporal networks (colored regions in Figs.~\ref{fig:I_i}(c) and \ref{fig:I_i}(d))
for intermediate values of $\taud$ ($10^2 \leq \taud \leq 10^4$ and $10^4 \leq \taud \leq 10^6$ for Conference and Email data sets, respectively).
In the original data sets, the nodes tend to have many events in bursty periods and be quiescent in other periods.
The randomization procedure eliminates bursty activity patterns.
Therefore, $v_m(t)$ can reach $\vthr$ in such a bursty period for the original but not randomized temporal networks if $\taud$ and $\vthr$ take intermediate values.
In the randomized data sets, $v_m(t)$ tends to decay faster than it grows,
although the number of events per node is the same between the original and randomized data. 

For Email data set, $I_m$ for the randomized data set (Fig.~\ref{fig:I_i}(d)) is larger
than that for the original data set (Fig.~\ref{fig:I_i}(c)) when $\taud$ is large and $\vthr$ is small.
This is mainly because the randomization increases the reachability ratio from initial seed $m$ to a large extent.
The reachability ratio from a node is defined as the fraction of nodes that we can reach from the node by tracing the events in the chronological order~\cite{Holme2005}.
If every event can elicit infection, which is the case when $\taud$ is large and $\vthr$ is small,
$I_m$ is approximated by the reachability ratio from node $m$.
The reachability ratio from node $m=3024$ in Email data set is equal to 0.7458 and 0.9981
for the original and randomized data sets, respectively. 
In contrast, the reachability ratio from node $m=55$ in Conference data set is equal to 0.9642 and 1
for the original and randomized data sets, respectively; the difference is smaller than in the case of Email data set.

\begin{figure}
\includegraphics[width=0.49\hsize,clip]{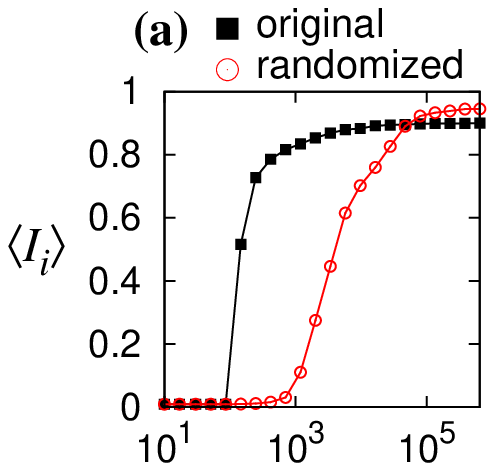}
\includegraphics[width=0.43\hsize,clip]{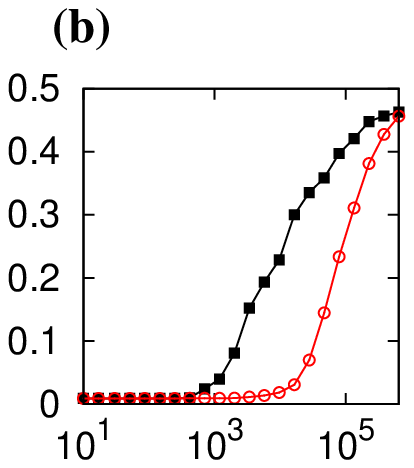}\\
\includegraphics[width=0.49\hsize,clip]{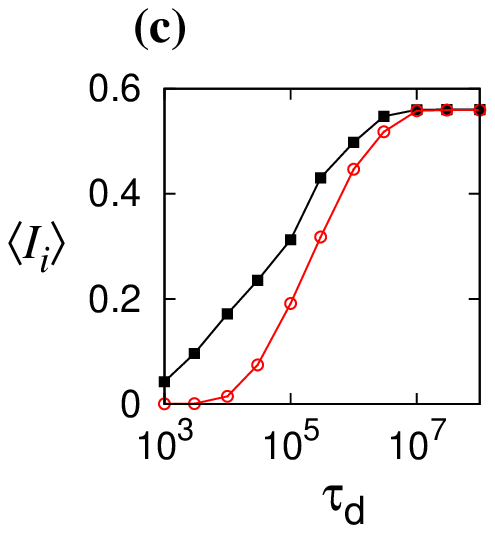}
\includegraphics[width=0.43\hsize,clip]{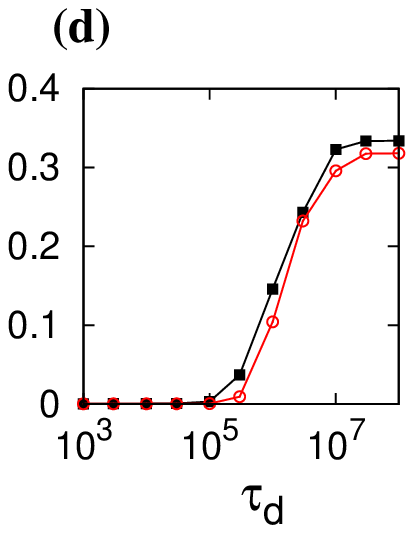}
\caption{Average final infection size $\langle I_i \rangle$ for (a, b) Conference and (c, d) Email data sets.
Squares and circles correspond to the original and randomized temporal networks, respectively.
We set (a) $\vthr=5$, (b) $\vthr=20$, (c) $\vthr=3$, and (d) $\vthr=10$. }
\label{fig:avg_I}
\end{figure}

In Fig.~\ref{fig:avg_I}, the average final infection size $\langle I_i \rangle$,
defined as the average of $I_i$ over all the nodes~$i$, is plotted as a function of $\taud$ for two values of $\vthr$ for each data set.
Figure \ref{fig:avg_I} indicates that $\langle I_i \rangle$ for the original temporal networks is larger than that for the randomized temporal networks for a broad range of $\taud$ for both data sets.

In the bond percolation on static networks, the probability that single bonds are open (independent of different bonds) is the sole parameter that determines the possibility that the entire network has a giant component~\cite{Newman2010}.
Motivated by this picture, we hypothesize that the results shown in Figs.~\ref{fig:I_i} and \ref{fig:avg_I} are largely explained by the bursty nature of events on single links. In other words, we speculate that the structure of the aggregated networks or correlation between event sequences on different links do not much influence the results.
To test the hypothesis, we separately examine the event sequence on each link.
For each link, \ie, node pair $(i,j)$ with at least one event,
$T_{j \to i}(t)$ is defined as the time required for node $i$ to be infected since node $j$ has been infected.
We emphasize that we do not consider influences from other nodes on $i$ in this analysis.
We take the time average of $T_{j \to i}(t)$, denoted by $\overline{T}_{j \to i}$, over $0 \leq t \leq t_{\max}$.
A problem with the time averaging is that $T_{j \to i}(t)$ is indefinite for sufficiently large $t$ because $i$ does not get infected by time $t_{\max}$.
Therefore, we adopt the boundary condition in which the first events between nodes $i$ and $j$ virtually replay after $t = t_{\max}$.
We denote the time of the first event between $i$ and $j$ by $t_1$.
If we temporarily set $T_{j \to i}(t_{\max}+t_1) = T_{j \to i}(t_1)$,
it takes at most $t_{\max} -t + t_1 + T_{j \to i}(t_1)$ for node $i$ starting with $v_i(t)=0$ to be infected from node $j$,
where $t_{\rm last} \leq t \leq t_{\max}$ and $t_{\rm last}$ is the last time before which $T_{j \to i}(t)$ is finite.
Therefore, we set $T_{j \to i}(t) = t_{\max} -t + t_1 + T_{j \to i}(t_1)$ for $t_{\rm last} \leq t \leq t_{\max}$.
This boundary condition is the same as that is used in Ref.~\cite{Pan2011} for defining the average temporal path length.
If $T_{j \to i}(t_1)$ is indefinite (\ie, infection never occurs between $i$ and $j$), $\overline{T}_{j \to i}$ is set to infinite. 
We define  denoted by $\langle 1 / \overline{T}_{j \to i} \rangle$ as the average of $1 / \overline{T}_{j \to i}$ over the 20\% links with the largest numbers of events,
because the majority of the links possesses a small number of events in both data sets.
This thresholding leaves 441 and 6,932 links for Conference and Email data sets, respectively.

\begin{figure}
\includegraphics[width=0.49\hsize,clip]{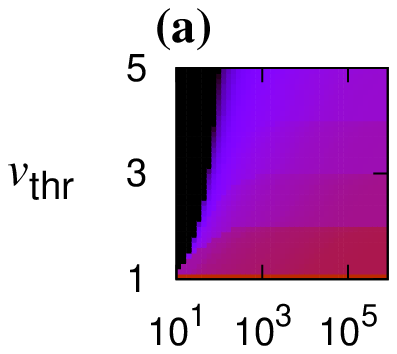}
\includegraphics[width=0.49\hsize,clip]{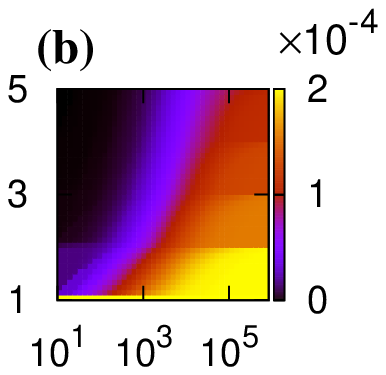}\\
\includegraphics[width=0.49\hsize,clip]{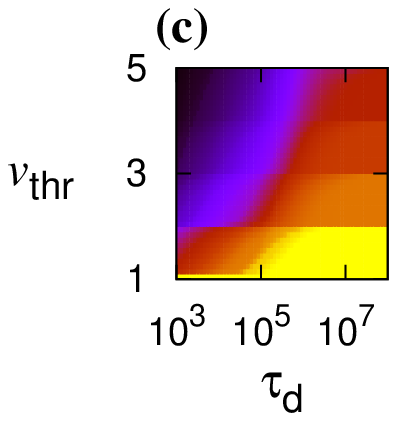}
\includegraphics[width=0.49\hsize,clip]{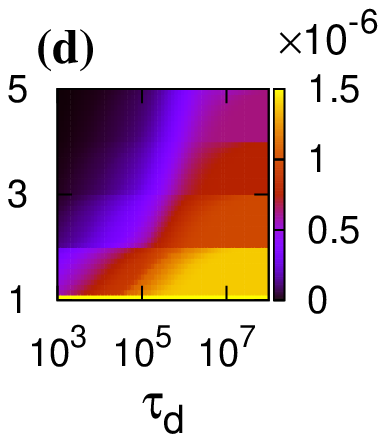}
\caption{Average single-link infection rate $\langle 1/\overline{T}_{j \to i} \rangle$ for (a), (b) Conference and (c), (d) Email data sets. (a), (c) Original temporal networks. (b), (d) Randomized temporal networks.}
\label{fig:firing_rate}
\end{figure}

$\langle 1 / \overline{T}_{j \to i} \rangle$ for the original and randomized temporal networks are shown for various $\taud$ and $\vthr$ values for Conference (Figs.~\ref{fig:firing_rate}(a) and \ref{fig:firing_rate}(b)) and Email (Figs.~\ref{fig:firing_rate}(c) and \ref{fig:firing_rate}(d)) data sets.
Because infection can be induced only through a single link in the present simulations,
we examined $\vthr$ values that are much smaller than those used in Figs.~\ref{fig:I_i} and \ref{fig:avg_I}.	
For both data sets, $\langle 1 / \overline{T}_{j \to i} \rangle$ for the original temporal networks (Figs.~\ref{fig:firing_rate}(a) and \ref{fig:firing_rate}(c)) is larger than that for the randomized networks (Figs.~\ref{fig:firing_rate}(b) and \ref{fig:firing_rate}(d)) for intermediate values of $\taud$ ($10^2 \leq \taud \leq 10^4$ and $10^4 \leq \taud \leq 10^6$ for Conference and Email data sets, respectively).
The behavior of $\langle 1 / \overline{T}_{j \to i} \rangle$ is consistent with the results of the network-based simulations (Figs.~\ref{fig:I_i} and \ref{fig:avg_I}).

%%%%%%%%%%%%%%%%%%%%%%%%%%%%%%%%%%%%%%%%%%%%%%%
\section{Conclusions}
We numerically simulated a variant of the deterministic threshold model on empirical temporal networks.
We found that the average final infection size for the empirical temporal networks is larger than those for the randomized temporal networks in a broad parameter region (Figs.~\ref{fig:I_i} and \ref{fig:avg_I}).
The bursty nature of the IEIs on single links has a sufficient explanatory power for the results of the network-based simulations (Fig.~\ref{fig:firing_rate}).
The burstiness promoted epidemic spreading when the decay exponent $\taud$ takes an intermediate value
($10^2 \leq \taud \leq10^4$ and $10^4 \leq \taud \leq 10^6$ (seconds) for Conference and Email data sets, respectively).
This range of $\taud$ may be practical because the influence of a pathogen that an individual has received may last for hours to days.

The finding that the burstiness facilitates the spreading also sheds light on a function of the redundant interaction events.
We previously found that about 80\% of the events are redundant in the sense that they affect little on bridging efficient temporal paths in Conference data set~\cite{Takaguchi2012}.
However, for the spreading dynamics in our model,
such redundant events play a crucial role in increasing $v_i(t)$ within bursty periods.

\section*{Acknowledgments}
The authors thank to the SocioPatterns collaboration (http://www.sociopatterns.org) for providing the data set.
This research is supported by the Aihara Project, the FIRST program from JSPS, initiated by CSTP.
T.T. acknowledges support provided through Grants-in-Aid for Scientific Research (No.~10J06281) from JSPS, Japan.
N.M. acknowledges support provided through Grants-in-Aid for Scientific Research (No.~23681033) from MEXT, Japan.
P.H. acknowledges support from the Swedish Research Council and the WCU program through the
National Research Foundation of Korea funded by the Ministry of Education, Science and Technology R31-2008-10029.

%%%%%%%%%%%%%%%%%%%%%%%%%%%%%%%%%%%%%%%%%%%%%%%
%% Reference %%

%%%%%%%%%%%%%%%%%%%%%%%%%%%%%%%%%%%%%%%%%%%%%%%
%% figures
\clearpage

\clearpage

\clearpage

\clearpage

%%%%%%%%%%%%%
%% table
\clearpage
\begin{table}
\centering
\caption{Statistics of the two data sets.}
\label{tab:datasets}
\begin{tabular}{|l|r|r|}\hline
 & \makebox[6em][c]{Conference} &  \makebox[6em][c]{Email}\\\hline
Number of nodes ($N$) & 113 & 3,188\\\hline
Number of events & 20,808 & 309,125\\\hline 
Recording period & 3 days & 83 days\\\hline
Time resolution & 20 sec & 1 sec\\\hline
\end{tabular}
\end{table}

\end{document}